\documentclass[prb,twocolumn,showpacs,preprintnumbers,amsmath,amssymb,superscriptaddress, bibnotes]{revtex4}

\usepackage{color}
\usepackage{graphicx}
\usepackage{dcolumn}
\usepackage{bm}
\usepackage{hyperref}

\begin{document}

\title{Stripe antiferromagnetic correlations in LaFeAsO$_{1-x}$F$_{x}$ probed by $^{75}$As NMR}

\author{Shunsaku~Kitagawa}
\email{shunsaku@scphys.kyoto-u.ac.jp}
\author{Yusuke~Nakai}
\author{Tetsuya~Iye}
\author{Kenji~Ishida}
\email{kishida@scphys.kyoto-u.ac.jp}
\affiliation{Department of Physics, Graduate School of Science, Kyoto University, Kyoto 606-8502, Japan}
\affiliation{TRIP, JST, Sanban-cho bldg., 5, Sanban-cho, Chiyoda, Tokyo 102-0075, Japan}

\author{Yoichi~Kamihara}
\affiliation{TRIP, JST, Sanban-cho bldg., 5, Sanban-cho, Chiyoda, Tokyo 102-0075, Japan}

\author{Masahiro~Hirano}
\affiliation{ERATO-SORST, Frontier Research Center, Tokyo Institute of Technology, JST, Yokohama 226-8503, Japan}
\affiliation{Frontier Research Center, Tokyo Institute of Technology, Yokohama 226-8503, Japan}

\author{Hideo~Hosono}
\affiliation{ERATO-SORST, Frontier Research Center, Tokyo Institute of Technology, JST, Yokohama 226-8503, Japan}
\affiliation{Frontier Research Center, Tokyo Institute of Technology, Yokohama 226-8503, Japan}
\affiliation{Materials and Structures Laboratory, Tokyo Institute of Technology, Yokohama 226-8503, Japan}

\date{\today}

\begin{abstract}
The anisotropy of the nuclear spin-lattice relaxation rate $1/T_{1}$ of $^{75}$As was investigated in the iron pnictide LaFeAs(O$_{1-x}$F$_{x}$) ($x = 0.07, 0.11,$ and 0.14) as well as LaFeAsO. While the temperature dependence of the normal-state $1/T_1T$ in the superconducting (SC) $x = 0.07$ is different from that in the SC $x = 0.11$, their anisotropy of $1/T_1$, $R \equiv (1/T_{1})_{H \parallel ab}/(1/T_{1})_{H \parallel c}$ in the normal state is almost the same ($\simeq$ 1.5). The observed anisotropy is ascribable to the presence of the local stripe correlations with $Q = (\pi, 0)$ or $(0, \pi)$. In contrast, $1/T_1$ is isotropic and $R$ is approximately 1 in the overdoped $x = 0.14$ sample, where superconductivity is almost suppressed. These results suggest that the presence of the local stripe correlations originating from the nesting between hole and electron Fermi surfaces is linked to high-$T_c$ superconductivity in iron pnictides.
\end{abstract}

\pacs{76.60.-k,	
74.25.-q, 
74.70.Xa 
}

\abovecaptionskip=-5pt
\belowcaptionskip=-10pt

\maketitle


Iron pnictide LaFeAs(O$_{1-x}$F$_{x}$) exhibits superconductivity in the vicinity of an antiferromagnetic (AF) phase,\cite{Y.Kamihara_JACS_2008} and thus the interplay between magnetism and superconductivity is one of the major issues to be clarified in the iron pnictide. In LaFeAs(O$_{1-x}$F$_{x}$), AF fluctuations due to nesting between hole and electron Fermi surfaces (FSs) originating from Fe3$d$ electrons were suggested from several theories,\cite{D.J.Singh_PRL_2008,I.I.Mazin_PRL_2008,K.Kuroki_PRL_2008} and stripe spin fluctuations with $\bm{Q}{\rm_{nesting}^{2D}}$~=~($\pi$,~0) and (0,~$\pi$) in the orthorhombic notation, which are identical to the nesting vectors, were observed by inelastic neutron scattering.\cite{M.Ishikado_JPSJ_2009,S.Wakimoto_arXiv_2009} 

We have investigated spin dynamics in LaFeAs(O$_{1-x}$F$_x$) through $^{75}$As and $^{139}$La nuclear magnetic resonance (NMR).\cite{Y.Nakai_JPSJ_2008,Y.Nakai_NJP_2009} 
In general, the nuclear spin-lattice relaxation rate divided by temperature $1/T_1T$, which is related to $q$-averaged low-energy dynamical electron-spin susceptibility, gives information on spin fluctuations. 
$1/T_1T$ in the undoped LaFeAsO increases on cooling due to strong AF fluctuations, and exhibits a pronounced peak at the AF ordering temperature $T_{\rm N} \simeq 140$~K. 
With F doping, the AF ordering and fluctuations are strongly suppressed and superconductivity is observed at $x \simeq 0.04$. 
Moreover, the appreciable AF fluctuations were not observed at $x = 0.11$ where superconducting (SC) transition temperature $T_{c}$ was reported to be maximum.\cite{Y.Nakai_JPSJ_2008,Y.Nakai_NJP_2009} 
These results suggest a weak correlation between low-energy AF fluctuations probed by NMR and $T_{c}$ values in LaFeAs(O$_{1-x}$F$_{x}$).

In contrast to LaFeAs(O$_{1-x}$F$_{x}$), low-energy AF fluctuations would play an important role in superconductivity in the ``122" and ``11" compounds since remarkable AF fluctuations have been observed in the maximum-$T_{c}$ compound in each system.\cite{Y.Nakai_PRB_2010,H.Fukazawa_JPSJ_2009,F.L.Ning_JPSJ_2009,T.Imai_PRL_2009,M.Yashima_JPSJ_2009} 
It is plausible that AF fluctuations can be masked by the decrease in $1/T_1T$ on cooling in LaFeAs(O$_{1-x}$F$_{x}$) due to band structure effects. 
Indeed, the recent theoretical studies indicate that the decrease in $1/T_1T$ on cooling observed in electron-doped iron pnictides can be interpreted by the characteristic band dispersion around the Fermi energy.\cite{H.Ikeda_JPSJ_2008,H.Ikeda_arXiv_2010} 
Therefore, one may consider that it is difficult for NMR measurements to probe whether the Fe spin fluctuations with $\bm{Q}$~=~($\pi$,~0) and (0,~$\pi$) are present or not in the SC LaFeAs(O$_{1-x}$F$_{x}$). 
However, it was pointed out that the anisotropy of $1/T_{1}$, [$R$ $\equiv (1/T_1)_{H \parallel ab}/(1/T_1)_{H \parallel c}$ : $(1/T_1)_i$ is measured in the field along $i$ direction] gives information on the local spin correlations due to the off diagonal terms in the hyperfine coupling tensor, and showed that the stripe AF correlations can be identified from NMR measurements in BaFe$_2$As$_2$ and SrFe$_2$As$_2$.\cite{K.Kitagawa_JPSJ_2009} 

Here, we report $1/T_{1}$ of $^{75}$As for LaFeAs(O$_{1-x}$F$_{x}$) ($x$ = 0, 0.07, 0.11, and 0.14) measured in magnetic fields parallel and perpendicular to the $c$ axis. On the basis of our results, we have uncovered the hidden relationship between stripe AF correlations and superconductivity in LaFeAs(O$_{1-x}$F$_{x}$). 

\begin{figure}[t]
\vspace*{-15pt}
\begin{center}
\includegraphics[width=9cm,clip]{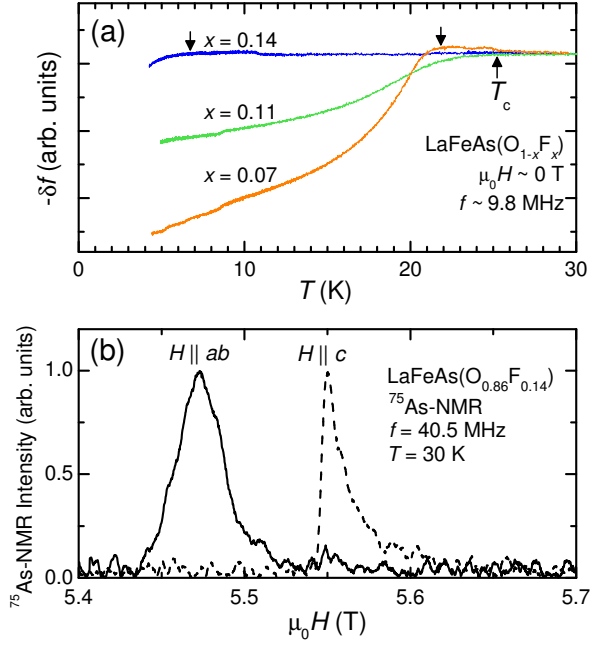}
\end{center}
\caption{(Color online) (a) Meissner signals at $x=0.07, 0.11,$ and 0.14 measured by an identical NMR coil. (b) Field-swept $^{75}$As-NMR spectra for uniaxially aligned LaFeAs(O$_{0.86}$F$_{0.14}$) in $H \parallel ab$ (solid line) and $H \parallel c$ (dashed line). 
}
\label{Fig.1}
\end{figure}

We performed $^{75}$As-NMR measurements in polycrystalline samples of LaFeAs(O$_{1-x}$F$_{x}$) ($x = 0.07, 0.11,$ and 0.14), which were used in our previous studies.\cite{Y.Nakai_JPSJ_2008,Y.Nakai_NJP_2009} 
$T_{c}$ of $x = 0.14$ (6.7~K) is lower than $T_{c}$ of $x = 0.07$ (21.8~K) and 0.11 (25.2~K), which were determined from the onset temperature of Meissner signal measured by a NMR coil as shown in Fig.~1 (a). Superconductivity at $x = 0.14$ is almost suppressed, since its Meissner signal is much smaller than that of $x = 0.07$ and 0.11, and any clear decrease in $1/T_1T$ below $T_c$ was not observed at $x = 0.14$.\cite{Y.Nakai_NJP_2009} All the samples were ground into powder, mixed with stycast 1266, and were rotated in the external magnetic field of 1.4~T while the stycast cures. Uniaxially aligned samples were thus prepared as shown in Fig. \ref{Fig.1} (b).\cite{B.L.Young_RSI_2002} 
$1/T_{1}$ was measured in $\mu_{0}H$~$\simeq$~9.89~T at 72.1~MHz for $H~\parallel~ab$ and in $\mu_{0}H$~$\simeq$~5.55~T at 40.5~MHz for $H~\parallel~c$. 
$1/T_{1}$ in the normal state was found to be $H$ independent within the measured field range.

\begin{figure}[t]
\vspace*{-27pt}
\begin{center}
\includegraphics[width=9cm,clip]{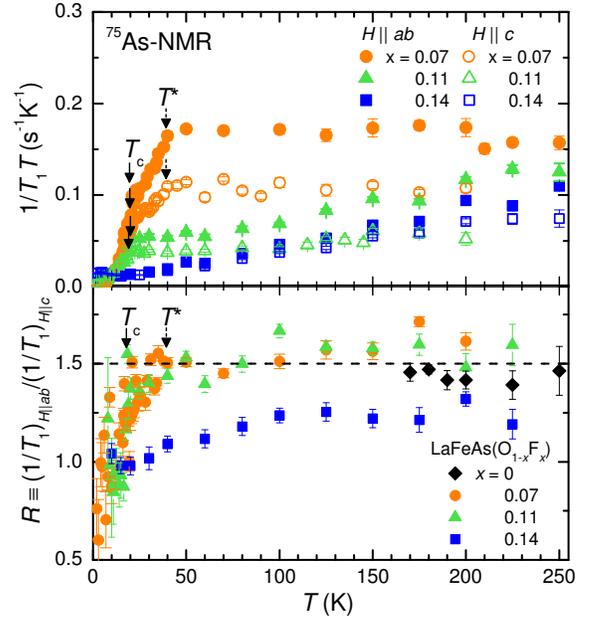}
\end{center}
\caption{(Color online) (Upper panel) $T$ dependence of $1/T_{1}T$ for $H \parallel ab$ and $H \parallel c$ in LaFeAs(O$_{1-x}$F$_{x}$). 
The solid (dashed) arrows represent $T_{c}$ ($T^{*}$). 
At $x = 0.07$ and 0.11, normal state $1/T_{1}T$ was anisotropic, while clear anisotropy in $1/T_{1}T$ was not observed at $x = 0.14$. (Lower panel) $T$ dependence of the anisotropy of $1/T_1$ [$R \equiv (1/T_{1})_{H \parallel ab}/(1/T_{1})_{H \parallel c}$]. $R$ in the normal state is $\simeq 1.5$ at $x =0.07$ and 0.11 as well as in LaFeAsO, whereas $R$ is $\simeq 1$ at $x = 0.14$. } 
\label{Fig.2}
\end{figure}

The upper panel of Fig.~\ref{Fig.2} shows the $T$ dependence of $1/T_{1}T$ in $H \parallel ab$ and $H \parallel c$ for LaFeAs(O$_{1-x}$F$_{x}$) ($x = 0.07, 0.11,$ and 0.14). 
Although $1/T_{1}T$ in $H \parallel ab$ was reported in our previous paper,\cite{Y.Nakai_JPSJ_2008,Y.Nakai_NJP_2009} $1/T_{1}T$ in $H \parallel c$ is a recent result, which can be obtained with uniaxially aligned samples.
As reported previously, $1/T_{1}T$ remains nearly constant down to $T^{*} \simeq$ 40~K and starts to decrease below $T^{*}$ and rapidly below $T_c$ at $x= 0.07$. $1/T_{1}T$ decreases from room temperature on cooling and approaches constant values near $T_c$ at $x = 0.11$ and 0.14. $1/T_{1}T$ at $x = 0.07$ and 0.11 are anisotropic in the normal state, while $1/T_{1}T$ at $x = 0.14$ is nearly isotropic. The anisotropy of $1/T_{1}$, [$R \equiv (1/T_{1})_{H \parallel ab}/(1/T_{1})_{H \parallel c}$] is plotted against temperature in the lower panel of Fig.~\ref{Fig.2}, along with $R$ in LaFeAsO. At $x = 0.07$ and 0.11, the $R$ in the normal state is $\simeq 1.5$ and decreases below $T^{*}$, 
while it is approximately 1.2 at 225 K and gradually decreases to 1 at $T_c$ for $x = 0.14$, indicative of nearly isotropic spin fluctuations at $T_c$. 
In the SC state, $R$ decreases to $\simeq 1$ and $\simeq 0.5$ at $x = 0.07$ and 0.11, respectively, which would be ascribed to the anisotropy of $H_{c2}$ and/or the vortex state $1/T_1$. 

In order to discuss spin correlations in the normal state, we calculate the anisotropy of $1/T_{1}$ for three types of spin correlations. We assume that hyperfine fields at the As site $\bm{H}\rm{_{hf}^{As}}$ are determined by the sum of the fields from the four nearest-neighbor Fe electron spins $\bm{S}$ 
\begin{align}
\bm{H}_{\rm{hf}}^{\rm{As}} = \sum_{i=1}^{4} \bm{B_{i}} \cdot \bm{S}_{i}
            = \Tilde{A} \bm{S},
\label{eq.1}
\end{align}
where $\bm{S}_{i}$ is the Fe electron spin at the $i$th Fe site, $\bm{B_{i}}$ is the hyperfine coupling tensor between the As nucleus and $i$th Fe site, and $\Tilde{A}$ is the hyperfine coupling tensor ascribed to the four nearest-neighbor Fe electron spins. 
Following the previous discussion,\cite{K.Kitagawa_JPSJ_2008} $\Tilde{A}$ can be described as follows in the orthorhombic notation:
\begin{align}
\Tilde{A} = \left(\begin{array}{ccc}
  A_{a} & C      & B_{1} \\
  C      & A_{b} & B_{2} \\
  B_{1}  & B_{2}  & A_{c} \\
  \end{array} \right).
\label{eq.2}
\end{align}
$A_{i}$ is a diagonal term for the $i$ direction ($i = a, b,$ and $c$), which is related to the Knight-shift components. $B_{1\{2\}}$ components are related with the stripe ($\pi$,~0) \{(0,~$\pi$)\} AF correlations and $C$ is related with the checkerboard ($\pi$,~$\pi$) AF correlations. 

In general, $1/T_{1}$ can be described in terms of fluctuating hyperfine fields perpendicular to the applied magnetic field parallel to the $z$ axis 
\begin{align}
\left(\frac{1}{T_{1}}\right)_{z} & = \frac{(\mu_{0} \gamma_{\rm{N}})^{2}}{2} \int^{\infty}_{- \infty} dt e^{i \omega_{\rm{res}} t} (\langle {H_{\rm{hf},x}(t), H_{\rm{hf},x}(0)} \rangle \notag \\
&+ \langle {H_{\rm{hf},y}(t), H_{\rm{hf,y}}(0)} \rangle) \notag\\
& = (\mu_{0} \gamma_{\rm{N}})^{2} (|H_{\rm{hf},x}(\omega_{\rm{res}})|^{2} + |H_{\rm{hf},y}(\omega_{\rm{res}})|^{2}), 
\label{eq.3}
\end{align}
where $|X(\omega)|^{2}$ denotes the power spectral density of a time-dependent random variable $X(t)$. 
Since ($\pi$,~0) and (0,~$\pi$) AF correlations cannot be distinguished in the tetragonal phase, we consider the three types of spin correlations: uncorrelated (UC) fluctuations centered at $q \simeq (0,0)$, stripe ($\pi$,~0), and checkerboard ($\pi$,~$\pi$) correlations. 
From Eqs.~(\ref{eq.1}) - (\ref{eq.3}), $1/T_{1}$ at the As site can be described as follows: 
\begin{align*}
\left(\begin{array}{c}
(1/T_{1})_{H \parallel a} \\
(1/T_{1})_{H \parallel b} \\
(1/T_{1})_{H \parallel c} \\
\end{array} \right)
&\propto \left(\begin{array}{c}
|A_{b}S_{b}(\omega_{\mathrm{res}})|^{2} + |A_{c}S_{c}(\omega_{\mathrm{res}})|^{2} \\
|A_{c}S_{c}(\omega_{\mathrm{res}})|^{2} + |A_{a}S_{a}(\omega_{\mathrm{res}})|^{2} \\
|A_{a}S_{a}(\omega_{\mathrm{res}})|^{2} + |A_{b}S_{b}(\omega_{\mathrm{res}})|^{2} \\
\end{array} \right)\\
&\hspace*{-75pt}\mathrm{for~UC~fluctuations}\notag\\
&\propto \left(\begin{array}{c}
|B_{1}S_{a}(\omega_{\mathrm{res}})|^{2}\\
|B_{1}S_{a}(\omega_{\mathrm{res}})|^{2} + |B_{1}S_{c}(\omega_{\mathrm{res}})|^{2}\\
|B_{1}S_{c}(\omega_{\mathrm{res}})|^{2}\\
\end{array} \right)\\
&\hspace*{-75pt}\mathrm{for~stripe~(\pi,~0)~and}\notag\\
&\propto \left(\begin{array}{c}
|C~S_{a}(\omega_{\mathrm{res}})|^{2}\\
|C~S_{b}(\omega_{\mathrm{res}})|^{2}\\
|C~S_{a}(\omega_{\mathrm{res}})|^{2} + |C~S_{b}(\omega_{\mathrm{res}})|^{2} \\
\end{array} \right)
\end{align*}
for~checkerboard~($\pi,~\pi$).
Assuming that fluctuating Fe spin components are isotropic in the spin space, 
$|S_{a}(\omega_{\mathrm{res}})|$~=~$|S_{b}(\omega_{\mathrm{res}})|$~=~$|S_{c}(\omega_{\mathrm{res}})|$, the $R \equiv (1/T_{1})_{H \parallel ab}/(1/T_{1})_{H \parallel c}$ is calculated as follows: 
\begin{align}
R = \begin{cases}
\frac{(A^{2}_{b} + A^{2}_{c} + A^{2}_{c} + A^{2}_{a})/2}{A^{2}_{a} + A^{2}_{b}} = \frac{A^{2}_{ab} + A^{2}_{c}}{2A^{2}_{ab}}= 0.5 + 0.5 \left(\frac{A_{c}}{A_{ab}}\right)^{2} \\ \hspace{70pt} \mathrm{for~UC},\\
\frac{(B^{2}_{1} + 2 B^{2}_{1})/2}{B^{2}_{1}} = 1.5 \hspace{12pt} \mathrm{for~(\pi,~0)},\\
\frac{(C^{2} + C^{2})/2}{2 C^{2}} = 0.5 \hspace{15pt} \mathrm{for~(\pi,~\pi)},
\end{cases}
\label{eq.4}
\end{align}
where $(1/T_{1})_{H \parallel ab} = \frac{(1/T_{1})_{H \parallel a} + (1/T_{1})_{H \parallel b}}{2}$ and $A_{ab}~\equiv~A_{a}~=~A_{b}$ in the tetragonal phase. 

\begin{figure}[b]
\vspace*{-15pt}
\begin{center}
\includegraphics[width=9.5cm,clip]{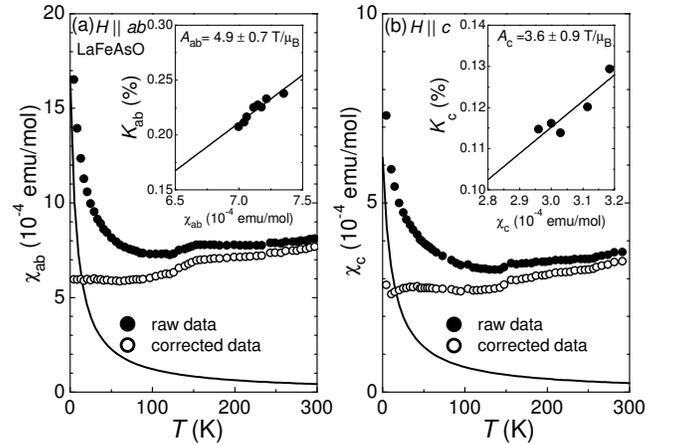}
\end{center}
\caption{(Main panel) $T$ dependence of the susceptibility in single-crystalline LaFeAsO in the field parallel to (a) the $ab$ plane ($\chi_{ab}$) and (b) the $c$ axis ($\chi_{c}$). The susceptibility data were obtained from Ref.~19. The CW behavior ascribed to approximately 2\% impurities was observed at low temperature. The $T$ dependence of $\chi_{ab}$ and $\chi_{c}$ was fitted to a $\frac{a}{T - \theta}~+~\chi_{0}$ formula. The fitting parameters are following: $a$~=~(131~$\pm$~2)~$\times$~$10^{-4}$~${\rm emu K/mol}$ and $\theta$~=~-~7.9~$\pm$~0.2~K for $\chi_{ab}$, and $a$~=~(74~$\pm$~2)~$\times$~$10^{-4}$~${\rm emu K/mol}$ and $\theta$~=~-~12~$\pm$~0.5~K for $\chi_c$. 
The corrected susceptibilities, which were used for the estimation of $A_i$ ($i$ = $ab$ and $c$), were derived by subtracting the CW term (solid line). (Inset) The Knight shift versus the corrected susceptibility in single-crystalline LaFeAsO. The solid lines represent linear fitting above 160~K. }
\label{Fig.3}
\end{figure}
Now, we consider the anisotropy of the diagonal $A_{i}$ terms in the hyperfine coupling tensor, which can be estimated based on the plot of the Knight shift against the bulk susceptibility. At present, since the bulk susceptibility measured in a single crystal was reported only for LaFeAsO,\cite{J.Q.Yang_APL_2009} the Knight shift perpendicular to the $c$ axis ($K_{ab}$) and along the $c$ axis ($K_{c}$) were measured in the uniaxially aligned LaFeAsO. $A_{ab}$ and $A_{c}$ in LaFeAsO were determined with the following relation, 
\begin{align}
K_{i} &=& \frac{\langle H_{\rm hf}\rangle}{H}+K_{\mathrm{orb},i}=\frac{A_{i} \langle S_i \rangle}{H_i}+K_{\mathrm{orb},i} \notag \\
&=& \frac{A_{i}}{N_{\rm{A}}\mu_{{\rm B}}} \chi_{i} + K_{\mathrm{orb},i}, (i = a, b,~\mathrm{and}~c), 
\label{eq.5}
\end{align}
where $\langle X \rangle$ is the time average of $X$, $N_{\rm{A}}$ is the Avogadro's number, $\mu_{\rm{B}}$ is the Bohr magneton, and $K_{\mathrm{orb}, i}$ is the orbital part of the Knight shift along the $i$ axis, which is generally $T$-independent. 
As seen in Fig.~\ref{Fig.3}, since the upturn (Curie tail) behavior due to approximately 2\% impurity was observed in the low-temperature bulk susceptibilities in both directions, $\chi_{ab}$ and $\chi_{c}$ were fitted to a Curie-Weiss (CW) function $\chi~=~\frac{a}{T - \theta}~+~\chi_{0}$, where $a$ and $\chi_{0}$ are $T$ independent, and $\theta$ is a Weiss temperature. To estimate the intrinsic bulk susceptibility in LaFeAsO, the Curie tail contribution was subtracted (see Fig.~\ref{Fig.3}). 
 
The insets of Fig.~\ref{Fig.3} show the plot of $K_{i}$ against the corrected susceptibility $\chi_{i}$, giving $A_{ab}~=~4.9~\mathrm{T}/\mu_{\mathrm{B}}$ and $A_{c}~=~3.6~\mathrm{T}/\mu_{\mathrm{B}}$ in LaFeAsO. 
The diagonal hyperfine-coupling terms of LaFeAsO are approximately two times larger than those of $A$Fe$_{2}$As$_{2}$ ($A$~=~Ba, Sr) (Refs.16 and 18) while the anisotropy of the diagonal terms, $A_{c}/A_{ab} \simeq 0.7$, is nearly the same among these compounds. The value of $B_{1}$, which is related to the stripe ($\pi$, 0) correlations, can be known from the internal field at the As site $H_{\rm int}$ in the AF state, since the ordered Fe moments pointing to the $a$ axis with ($\pi$,~0) correlations give rise to the internal field along the $c$ axis at the As site. 
In LaFeAsO, it was reported that $H_{{\rm int}}$~=~1.60~T (Ref.20) and $\mu$~=~0.36$\mu_{\rm B}$.\cite{Cruz_Nature_2008} These values lead to $B_{1}~=~4.4~\mathrm{T}/\mu_{\mathrm{B}}$ in LaFeAsO, which is also nearly two times larger than $B_{1}~=~1.72~\mathrm{T}/\mu_{\mathrm{B}}$ in BaFe$_2$As$_2$ and $B_{1}~=~2.08~\mathrm{T}/\mu_{\mathrm{B}}$ in SrFe$_2$As$_2$.  
It seems that the value of $A_{c}/A_{ab}$ does not change largely in iron pnictides, although the values of the hyperfine couplings in LaFeAsO are nearly twice larger than those in $A$Fe$_2$As$_2$ ($A$ = Ba and Sr).

The observed $A_{c}/A_{ab}$ value ($\simeq 0.7$) in LaFeAsO, which is smaller than 1, cannot make $R$ to be larger than 1 as expected from Eq. \eqref{eq.4}. 
Therefore, the observed $R \simeq$ 1.5 in LaFeAsO above the structural-transition temperature $T_{\rm S}$ is consistently interpreted by the presence of the local ($\pi$,~0)~AF correlations, which were detected by the neutron-scattering measurement.\cite{M.Ishikado_JPSJ_2009} 
Similarly, $R \simeq 1.5$ observed at $x = 0.07$ and 0.11 also suggests the presence of the local stripe correlations, although any pronounced development of spin fluctuations was not detected from the temperature dependence of $1/T_1T$. 

In contrast, the anisotropy of 1/$T_{1}$ becomes approximately 1.0 at $x = 0.14$, indicating that the stripe $(\pi,0)$ correlations are weak or absent. The local stripe correlations can be induced by the nesting between hole and electron FSs. Therefore, the weakness or absence of the stripe correlations at $x = 0.14$ is ascribable to a worse nesting condition, since the hole FSs become smaller with F doping and would disappear at a critical concentration. The concentration of $x \simeq$ 0.14 could correspond to a critical one, since $R$ changes abruptly from $x = 0.11$. 


Recently, inelastic neutron scattering (INS) measurements on LaFeAs(O$_{1-x}$F$_{x}$) were reported.\cite{S.Wakimoto_arXiv_2009} They found that the two-dimensional stripe AF fluctuations, whose strength is comparable to those of LaFeAsO, are present in the SC $x = 0.057$ and 0.087 samples, but absent in the overdoped $x = 0.158$ sample where superconductivity is almost suppressed as in our $x = 0.14$ sample. Their results suggest that the spin fluctuations due to the FS nesting are indispensable to superconductivity. 
Our present NMR results of the anisotropy of $1/T_1$ seem to be quite consistent with their INS results. 
However, the weak coupling between the AF fluctuations and superconductivity, which is suggested from the strong suppression of $1/T_1T$ by F doping, might be inconsistent with their INS results. 
We consider that this discrepancy between the INS and NMR results originates from the characteristic energy of the stripe AF fluctuations; 
NMR measurements are sensitive to low energy (mK order) spin dynamics, but cannot detect high energy spin dynamics probed with INS measurements (K order). 
Such different behavior between low-energy and high-energy spin dynamics is also pointed out from the theoretical calculations.\cite{H.Ikeda_JPSJ_2008,H.Ikeda_arXiv_2010} 
According to these calculations, the imaginary part of the local spin susceptibility, Im$\chi^{\rm s}_{\rm loc}$ has a peak and increases on cooling at high-energy. 
In contrast, such an increase is not observed and Im$\chi^{\rm s}_{\rm loc}$ is suppressed by electron doping at low energy. 
Thus, elucidating the energy dependence of the stripe fluctuations and their relationship with superconductivity would be crucial for further understanding.  

Quite recently, $1/T_1T$ of $^{75}$As measured with $H\parallel ab$ and $H\parallel c$ has been reported in Ba(Fe$_{1-x}$Co$_{x}$)$_{2}$As$_{2}$ by Ning $et~ al.$\cite{F.L.Ning_JPSJ_2009,F.L.Ning_PRL_2010,F.L.Ning_arXiv_2009} By analyzing their data, $R$ is larger than 1.4 at $T_{c}$ in superconducting $x < 0.12$ samples, but $R$ is 1 in a whole temperature region in the overdoped $x = 0.26$ sample where superconductivity disappears. This result also gives evidence for the relationship between stripe correlations and superconductivity.

In summary, the anisotropy of the normal state $1/T_{1}$ of $^{75}$As, $R$, in LaFeAs(O$_{1-x}$F$_{x}$) was found to be $\simeq$ 1.5 at $x$ = 0.07 and 0.11, although their temperature dependences of the normal state $1/T_1T$ are quite different. 
By contrast, $R$ is $\simeq 1$ at $x =0.14$. On the basis of the simple model in which the off-diagonal terms in the hyperfine coupling tensor dominate $1/T_1$, our experimental results suggest the presence of the local stripe AF correlations originating from the nesting between the hole and electron FSs in the SC $x = 0.07$ and 0.11 samples, while such correlations are weak or absent at $x = 0.14$, where superconductivity is suppressed. 
These results suggest that the presence of the local stripe correlations originating from the nesting is important for the occurrence of high-$T_c$ superconductivity in the iron pnictides.

We are grateful to S. Yonezawa, H. Ikeda, and Y. Maeno for valuable discussions. 
This work is supported by a Grant-in-Aid for the Global COE Program ``The Next Generation of Physics, Spun from Universality and Emergence" from the Ministry of Education, Culture, Sports, Science, and Technology of Japan (MEXT), and for Scientific Research from the Japan Society for the Promotion of Science (JSPS). 


\end{document}